\documentclass[aps,twocolumn,showpacs,prb,floatfix,a4paper,superscriptaddress]{revtex4}

\usepackage{graphics,amsmath,amssymb}

\usepackage{graphicx}
\usepackage{color}
\usepackage{epstopdf}

\newcommand{\ivo}{In$_2$VO$_5$}

\newlength{\figwidth}

\divide\figwidth by 2

\graphicspath{{./}{Figures/}}

\begin{document}


\title{Insulator to semiconductor transition and magnetic properties of the one-dimensional S = 1/2 system In$_2$VO$_5$}

\author{A. M\"oller}\author{T. Taetz}
\affiliation{Institut f\"ur Anorganische Chemie, Universit\"at zu
K\"oln, Greinstr. 6,  50939 K\"oln, Germany}

\author{N. Hollmann}\author{J.A. Mydosh}
\affiliation{ II. Physikalisches Institut, Universit\"at zu K\"oln,
Z\"ulpicher Str. 77,  50937 K\"oln, Germany}

\author{V. Kataev}

\affiliation{Institute for Solid State and Materials Research IFW Dresden,
P.O. Box 270116, D-01171 Dresden, Germany}

\affiliation{Kazan Physical Technical Institute of the Russian Academy of
Sciences, 420029 Kazan, Russia}

\author{M. Yehia}
\affiliation{Institute for Solid State and Materials Research IFW Dresden,
P.O. Box 270116, D-01171 Dresden, Germany}

\author{E. Vavilova}
\affiliation{Institute for Solid State and Materials Research IFW Dresden,
P.O. Box 270116, D-01171 Dresden, Germany}

\affiliation{Kazan Physical Technical Institute of the Russian Academy of
Sciences, 420029 Kazan, Russia}

\author{B. B\"uchner}
\affiliation{Institute for Solid State and Materials Research IFW Dresden,
P.O. Box 270116, D-01171 Dresden, Germany}



\date{\today}
\begin{abstract}

We report structural, magnetization, electrical resistivity and
nuclear- and electron spin resonance data of the complex transition
metal oxide In$_2$VO$_5$ in which structurally well-defined V-O
chains are realized. An itinerant character of the vanadium
$d$-electrons and ferromagnetic correlations, revealed at high
temperatures, are contrasted with the insulating behavior and
predominantly antiferromagnetic exchange between the localized
V$^{4+}$ $S = 1/2$-magnetic moments which develop below a certain
characteristic temperature $T^*\sim 120$\,K. Eventually the compound
exhibits short-range magnetic order at $T_{\rm SRO}\approx 20$\,K.
We attribute this crossover occurring around $T^*$ to the unusual
anisotropic thermal contraction of the lattice which changes
significantly the overlap integrals and the character of magnetic
intra- and interchain interactions.

\end{abstract}


\maketitle

\section{Introduction}
There has been considerable interest in vanadium oxides over several
decades, due to their diverse structural chemistry and
unconventional physical properties. In particular metal-insulator
transitions occur in several binary vanadium-oxides as prominent
examples \cite{ueda2}. While studies of spin-ladder and spin-chain
compounds were mostly concentrated on cuprate compounds like
KCuBr$_3$ and TlCuCl$_3$ \cite{takatsu}, Sr$_2$CuO$_3$ and SrCuO$_2$
\cite{motoyama} or LaCuO$_{2.5}$ \cite{hiroi}, quasi low-dimensional
vanadium systems such as (VO)$_2$P$_2$O$_7$ \cite{johnston} or
A$_x$V$_2$O$_5$ (A = alkaline or alkaline-earth metal)
\cite{isobe,ueda} have experienced recent attention, due to their
variety of interesting quantum spin phenomena
\cite{ueda,dagotto1,dagotto2}. Spin ladders \cite{Vasilev} consist
of two (or more) parallel chains, often referred to as "legs", with
an intrachain exchange interaction parameter $J$. Interchain
interactions along the "rungs" are then represented by the parameter
$J^{\prime}$. In most cases with an even number of legs the smallest
magnetic entity is then given by a rectangle or exclusively by the
"rungs" which are equivalent to dimers. These systems are
representatives for a spin-gap.

In this paper we present a complex vanadium oxide, In$_2$VO$_5$, as
an experimental realization  of a one-dimensional two-leg spin-1/2
ladder with "sheared" legs, the so called zig-zag chain, where the
onset of antiferromagnetic (AFM) exchange along and between the legs
can be attributed to the magnetic properties of a triangle as the
smallest magnetic entity, whereby frustration phenomena might occur.
Although In$_2$VO$_5$ has been first synthesized long ago
\cite{senegas}, to our knowledge its electronic and magnetic
properties have not been explored so far. Here we report the crystal
structure, magnetization, resistivity, nuclear magnetic (NMR) and
electron spin resonance (ESR) measurements of In$_2$VO$_5$. We
observe a relationship between structural, charge and spin degrees
of freedom in this compound. In particular, experimental data give
evidence that an unusual anisotropic thermal contraction yields
localization of the itinerant $d$-states with decreasing temperature
below $\sim 120$\,K that concomitantly results in a crossover from
ferromagnetic (FM) to antiferromagnetic (AFM) exchange interaction
between the vanadium spins which eventually short-range order below
$\sim 20$\,K.

\section{Experimental}
Single crystal and powder samples for the present study were
prepared as described by Senegas \cite{senegas}. In$_2$O$_3$
(Chempur, 99.9\%) and VO$_2$ (Chempur, 99.5\%) were dried before
grounding. The reaction mixture of typically 1.3 g was then
encapsulated under a vacuum in a silica ampoule. The sealed ampoule
was heated directly to 1323\,K. After three days the furnace cooled
down to room temperature at a rate of 5\,K/h. Small black single
crystals of In$_2$VO$_5$ were obtained, selected under a microscope
and prepared for single crystal x-ray diffraction. The crystals were
of needle shape and typically 0.1$\times$0.1$\times$0.3 mm in size.
Single crystal x-ray diffraction was performed at 293\,K and 100\,K
on an IPDS II diffractometer (Stoe\&Cie). For powder x-ray
diffraction (Cu-K$_{\alpha}$ radiation) a Siemens D5000
diffractometer equipped with a self-assembled helium cryostat for
low temperature measurements was used. Lattice constants were
derived from a LeBail fitting routine using the program FullProf
\cite{fullprof}. Magnetization measurements were carried out on a
PPMS vibrating sample magnetometer (Quantum Design). The electrical
resistivity (standard four-point method) was measured on a pressed
and then at 1073\,K (12\,h) sintered powder pellet. The purity of
the sample after sintering was checked again by powder x-ray
diffraction. Absorption spectroscopy was performed on standard PE
(FIR) and KBr (MIR) pellets at room temperature using a Bruker
IFS-66v/s spectrometer. With KBr diluted powder pellets were
measured on a Cary 05E (Varian) spectrometer in the NIR-Vis-UV range
at temperatures from 293 to 20\,K. ESR measurements were carried out
at a frequency of $\nu$ = 9.5\,GHz with a standard X-Band Bruker EMX
Spectrometer and also in a frequency range of 140 to 360\,GHz with a
home made tunable high-field ESR spectrometer on the basis of the
Millimeterwave Vector Network Analyzer from AB Millimetre, Paris,
and in a 17\,T superconducting magnetocryostat from Oxford
Instruments Ltd. (see Ref. \cite{Golze06} for details). $^{51}$V NMR
data were recorded on a Tecmag pulse solid-state NMR spectrometer in
a magnetic field of 9.2\,T in a temperature range 4.2 - 200\,K. The
NMR spectra were acquired by a point-by-point magnetic field
sweeping. The transversal and the longitudinal relaxation times
$T_2$ and $T_1$ have been measured at the frequency of the maximum
intensity of the spectrum. $T_2$ has been determined from the Hahn
spin echo decay. A method of stimulated echo has been used to
measure $T_1$.

\section{Crystal structure}
\label{crystal-structure}

\begin{figure}
\label{struct}
\begin{center}
\includegraphics*[angle=0,width=0.9\columnwidth]{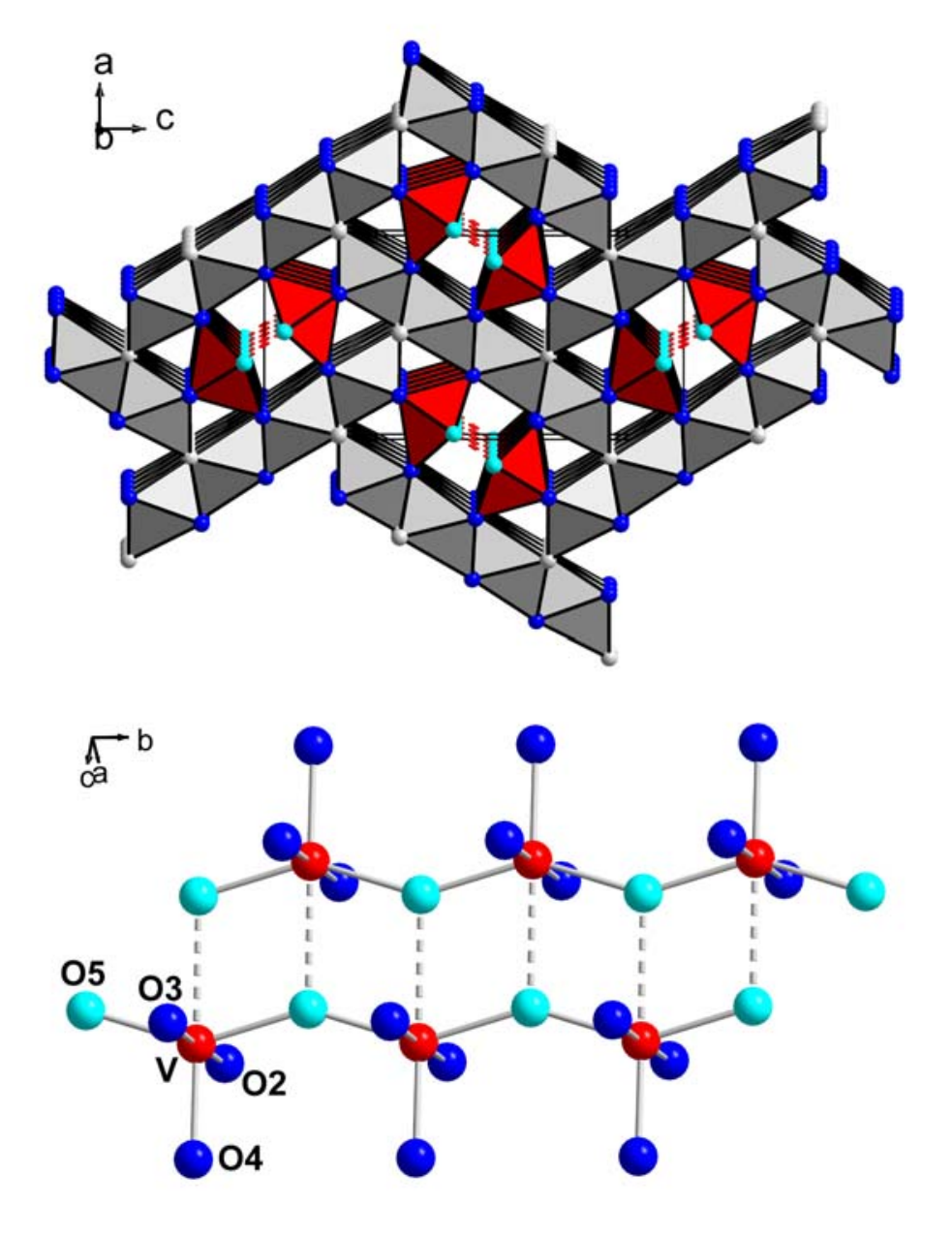}
\caption{Perspective view of the crystal structure of In$_2$VO$_5$ with
[InO$_6$] and [VO$_5$] polyhedra illustrated in grey and red,
respectively, O1 (white), O2-O4 (blue) and O5 (cyan). Below, the
structural feature of a zig-zag chain-type, $^1_{\infty}
$[\{O$_3$V(O5)$_{3/3}$\}$_2$], is depicted. The larger V-O5 distances are
represented by dotted lines. (Colors online)\label{view_b}}
\end{center}
\end{figure}

In 1975 Senegas \textit{et al.} \cite{senegas} reported the
synthesis and crystal structure of In$_2$VO$_5$ at room temperature.
We have re-determined the crystal structure at 293\,K and 100\,K see
Table~\ref{struc1}. The previous structure solution is confirmed by
our data \cite{icsd-invo}. Furthermore, no structural phase
transition has been observed from single crystal data down to 100 K.
Therefore, we limit the structure description to the important
structural features for understanding the physical properties and
refer to Senegas \textit{et al.} \cite{senegas} for a more detailed
discussion of the interatomic distances and angles.
Fig.~\ref{view_b} gives a perspective view of the crystal structure
of In$_2$VO$_5$. A three-dimensional network of [InO$_6$] octahedral
units is present. This constitutes of $^1_{\infty}$[In$_4$O$_{10}$]
slabs of edge sharing entities which are connected via corner
oxygens O1 with identical slabs, giving rise to a
$^3_{\infty}$[In$_4$O$_8$] framework. It should be noted that the
coordination sphere of O1 consists only of four In atoms in a
tetrahedral fashion with three In belonging to a common and one to
an independent slab. Thereby, a nonmagnetic anionic framework with
one-dimensional cavities along [010] is formed. The magnetic cations
(V$^{4+}$, $3d^1$, $S = 1/2$) are connected to the network via
oxygens O2, O3 and O4. Two O5 complement the coordination sphere of
each V, which can be described as a distorted trigonal bipyramid.
These entities are linked exclusively via O5 to zig-zag chains,
$^1_{\infty}$[O$_3$V(O5)$_{2/2}$], in [010]. The second coordination
sphere increases the C.N.(V) from 5 to 6 and the (sheared)
ladder-type structural feature is accomplished:
$^1_{\infty}$[\{O$_3$V(O5)$_{3/3}$\}$_2$]. The prominent structural
difference between 100\,K and 293\,K is the decreasing distance of
V-O5 connecting two legs by 1.5\,pm, marked by dotted lines in
Fig.~\ref{view_b}.

\begin{figure}
\begin{center}
\includegraphics*[angle=0,width=\columnwidth]{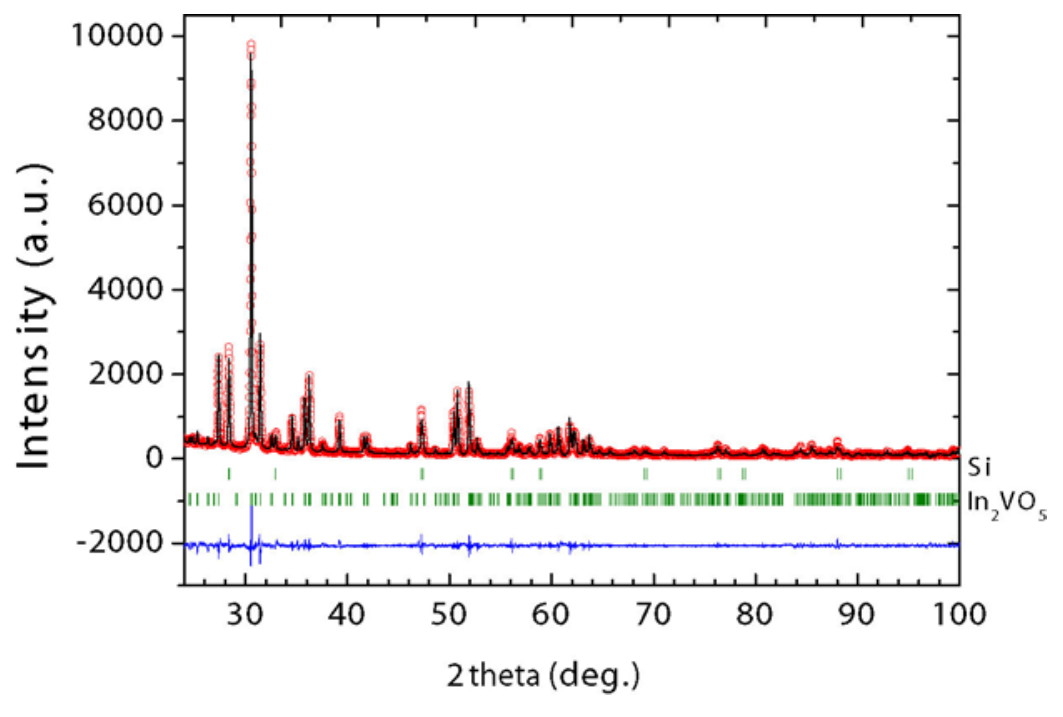}
\caption{Rietveld refinement of a x-ray powder diffraction pattern of
In$_2$VO$_5$ at 290 K. Si has been used as an internal standard. Lattice
parameters at room temperature as derived from Rietveld refinement are:
$a$ = 7.2265(9) \AA,~ $b$ = 3.4662(5) \AA,~ $c$ = 14.825(2) \AA.
Refinement parameters: $\chi^2$ = 2.96, R$_{\text{B}}$ = 1.39. (Colors
online)\label{diff}}
\end{center}
\end{figure}

\begin{figure}
\includegraphics*[angle=0,width=\columnwidth]{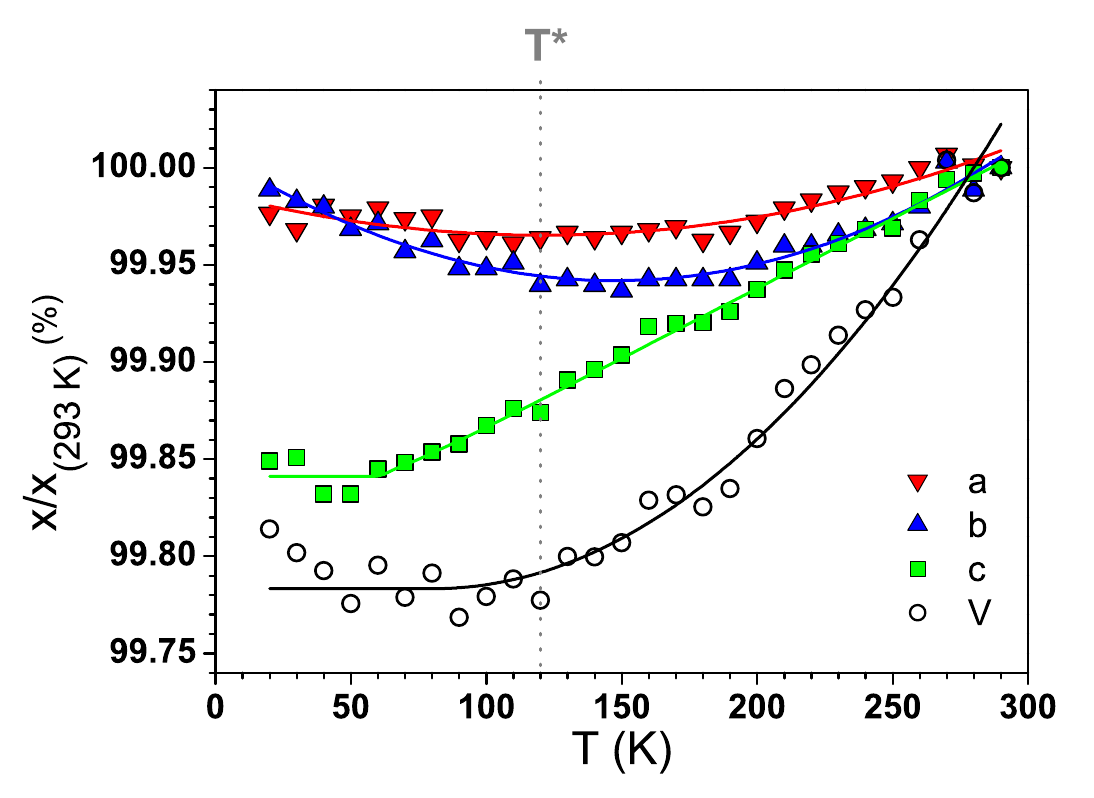}
\caption{Temperature dependence of lattice parameters and cell volume of
In$_2$VO$_5$ as determined from x-ray powder diffraction (lines are a
guideline for the principal changes). The anisotropic temperature
dependence sets in below $\sim$ 150\,K. For T $<$ T* $\sim$ 120\,K a
slight increase of the $a$ and $b$ axes occurs. (Colors
online)\label{lattice}}
\end{figure}

A typical refinement of x-ray powder data from which the lattice
parameters were extracted, can be seen in Fig.~\ref{diff}. The evolution
of the cell axes and the cell volume is shown in Fig.~\ref{lattice}. The
cell volume decreases roughly linearly with temperature down to 120\,K and
remains constant within standard deviations at temperatures below. The
overall decrease in cell volume from 290\,K to 120 K is $\sim$ 0.25 \%.

We found no indication of a structural transition down to 20\,K from
powder x-ray diffraction data, but rather a continuous change from
an isotropic decrease of the lattice constants to an anisotropic
behavior at $\sim$ 150\,K. Below 200\,K the decrease of the $a$ and
$b$ parameters levels off and turns into a slight increase by
further lowering the temperature from $T^* \sim 120$\,K to 20\,K.
This is in contrast to the behavior of the $c$ parameter, which
decreases almost linearly down to 50\,K. It should be noted that the
absolute values of the lattice parameters do not alter with
temperature by more than 0.2\% relative to 290\,K.

As will be shown below, the important consequence of the anisotropic
temperature dependence of the lattice parameters for the magnetism
of In$_2$VO$_5$ is that below a certain temperature $T^*$ the V-V
distance along [010] begins to increase slightly or stays almost
constant whereas it decreases for two adjacent chains.

\begin{table}
\caption{Lattice parameters and selected interatomic distances in pm
for In$_2$VO$_5$ (single crystal data). \label{struc1}}
\begin{ruledtabular}
\begin{tabular}{r c c c}
          & space group & P\textit{nma}&Z = 4      \\
          & 293 K       &        & 100 K       \\
$a$      & 725.1(1)    &        & 724.6(1)    \\
$b$      & 346.90(7)   &        & 346.68(7)   \\
$c$      & 1487.7(3)   &        & 1485.9(3)    \\
          &             &        &             \\
In -  O   & 210.0(3) - 224.0(2)&       & 210.1(4) - 223.8(2) \\
V - O4    &    178.1(4) &         &  178.2(4)    \\
O5$^i$    &  181.2(1)   &         &   181.1(1)   \\
O5$^{ii}$ &  181.2(1)   &         &   181.1(1)      \\
O2        &  201.9(4)   &         &  201.6(4)      \\
O3$^{iii}$ & 204.4(4)   &         &   203.9(4)      \\
O5$^{iii}$ & 226.9(4)   &         &  225.4(5)     \\
V - V      & 328.6(1) $^{a)}$  &     & 327.6(1) $^{a)}$              \\
           & 346.9(1) $^{b)}$ &        & 346.7(1) $^{b)}$             \\
\multicolumn{4}{l}{$^i$ 1-x, 0.5+y, 1-z  ~$^{ii}$ 1-x, y-0.5, 1-z
~$^{iii}$ x-1, y, z}\\
\multicolumn{4}{l}{$^{a)}$ "rungs" $^{b)}$  "legs"}\\
\end{tabular}
\end{ruledtabular}
\end{table}

\section{Magnetization measurements}
\label{magnetization}

\begin{figure}
\includegraphics*[angle=0,width=\columnwidth]{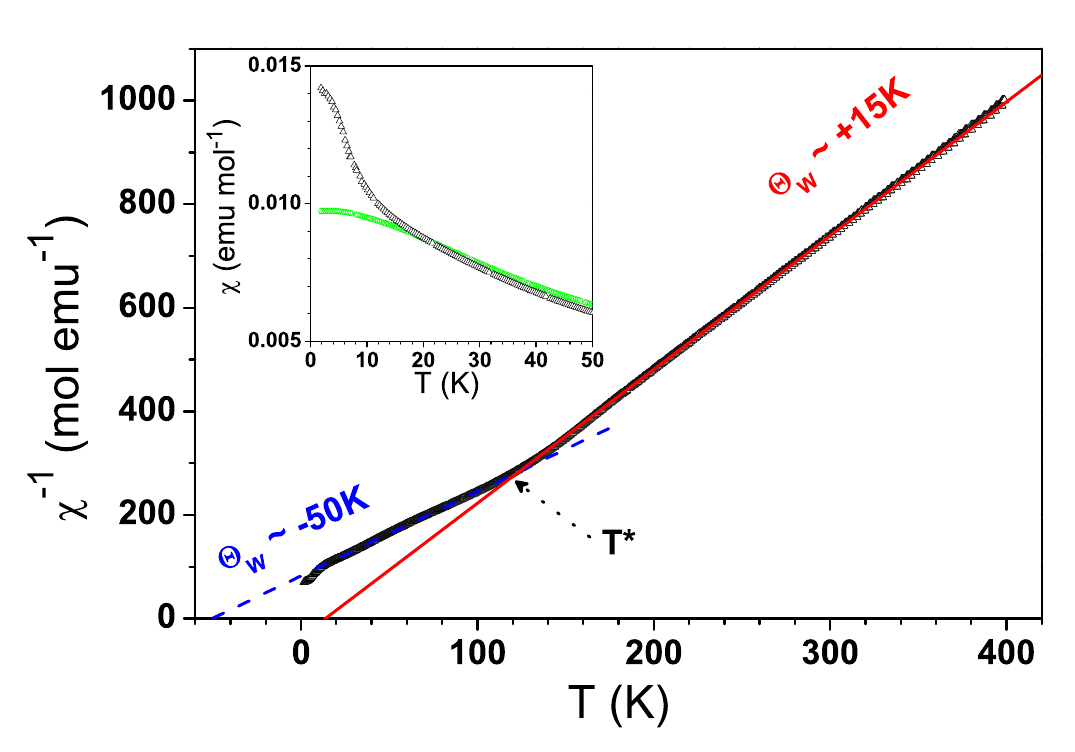}
\caption{Inverse susceptibility $\chi^{-1}(T)$ of In$_2$VO$_5$ at 1\,T
with linear fits according to the Curie-Weiss law for dominant ferro-
(red) and antiferro- (blue, dashed) magnetic interaction above and below
T*, respectively. The insert shows the static susceptibility $\chi(T)$ for
1\,T (black) and 14\,T (green). (Colors online)\label{chi}}
\end{figure}

The static susceptibility $\chi(T)$ of In$_2$VO$_5$ was measured on
a powder sample in external fields of 1\,T in field cooled (FC) and
zero field cooled (ZFC) modes and at 14\,T in a ZFC mode. The
results are shown in Fig.~\ref{chi}.

The inverse susceptibility $\chi^{-1}(T)$ exhibits a broad kink around a
characteristic temperature $T^*\sim 120$\,K which separates two linear
regimes with different slopes of  $\chi^{-1}(T)$. The fit of the
experimental data above $T^*$ to the Curie-Weiss law $\chi =
C/(T-\Theta_w)$ yields the Curie constant $C = 0.385$\,emu/molK and a
positive Weiss temperature $\Theta_w \sim + 15$\,K that indicates
ferromagnetic  (FM) interactions between the V spins. A similar fit at
$T<T^*$ yields $C = 0.615$\,emu/molK and a negative $\Theta_w \sim -
50$\,K implying the net AFM exchange between the V spins. The value of $C$
at $T>T^*$ corresponds approximately to the Curie constant of uncorrelated
paramagnetic $S=1/2$ moments with the $g$-factor $g=2$. An appreciable
increase of $C$ for $T<T^*$ indicates the significance of the magnetic
correlations in the low temperature regime which will be discussed below.

The FC and ZFC measurements of $\chi(T)$ in a field of 1\,T reveal no
differences over the whole temperature range. The ZFC curves for the two
fields 1\,T and 14\,T are also similar within the experimental error
except the low temperature range $T<15$\,K where the susceptibility
measured at 14\,T is smaller as compared to the 1\,T data (see inset of
Fig.~\ref{chi}).

\begin{figure}
\includegraphics*[angle=0,width=0.95\columnwidth]{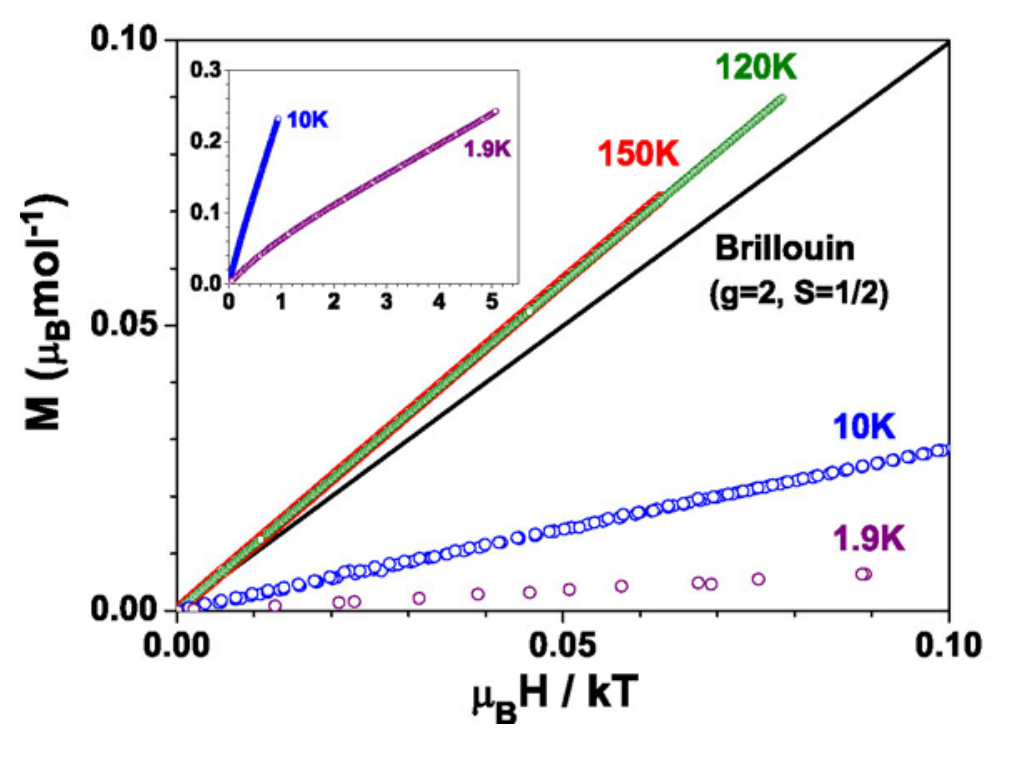}
\caption{Magnetization measurements at different temperatures for
In$_2$VO$_5$. The Brillouin function is given for a $S = 1/2$ system with
$g = 2$ (line). The inset shows the data for 10 K and 1.9 K in the high
field range. (Colors online)\label{mag2}}
\end{figure}

Measurements of the field dependence of the magnetization $M(H)$ at
several selected temperatures are shown in Fig.~\ref{mag2}. $M(H)$
curves reveal no hysteresis and are linear in the accessible field
range $H\leq 14$\,T except at the lowest temperature of 1.9\,K.  The
Brillouin function of noninteracting moments ($S=1/2$, $g = 2$) is
shown for comparison. One notices that compared to the Brillouin
function the experimental $M(H)$-curves exhibit larger slopes for
$T\geq 120$\,K whereas the slopes are strongly reduced at low
temperatures and no tendency to the saturation even at the highest
applied field can be seen. This evidences, similarly to the
$\chi(T)$ data, a crossover from the FM- to the AFM type of
interactions between the vanadium spins with decreasing temperature.
A nonlinear initial increase of $M(H)$ curve in small fields at
1.9\,K (see inset of Fig.~\ref{mag2}) suggests the occurrence of
$\sim 2$\,\% of impurity spins which are saturated in a field of a
few Tesla at low $T$s. That would explain the smaller low-$T$ values
of $\chi$ measured at $H = 14$\,T.

We note that although our static magnetic data reveal appreciable
magnetic correlations  no evidence for the occurrence of the
long-range magnetic order in In$_2$VO$_5$ can be found in the
studied temperature range \cite{sh}.

\section{Electrical resistivity and spectroscopic
characterization}
\label{resistivity}

\begin{figure}
\includegraphics*[angle=0,width=0.95\columnwidth]{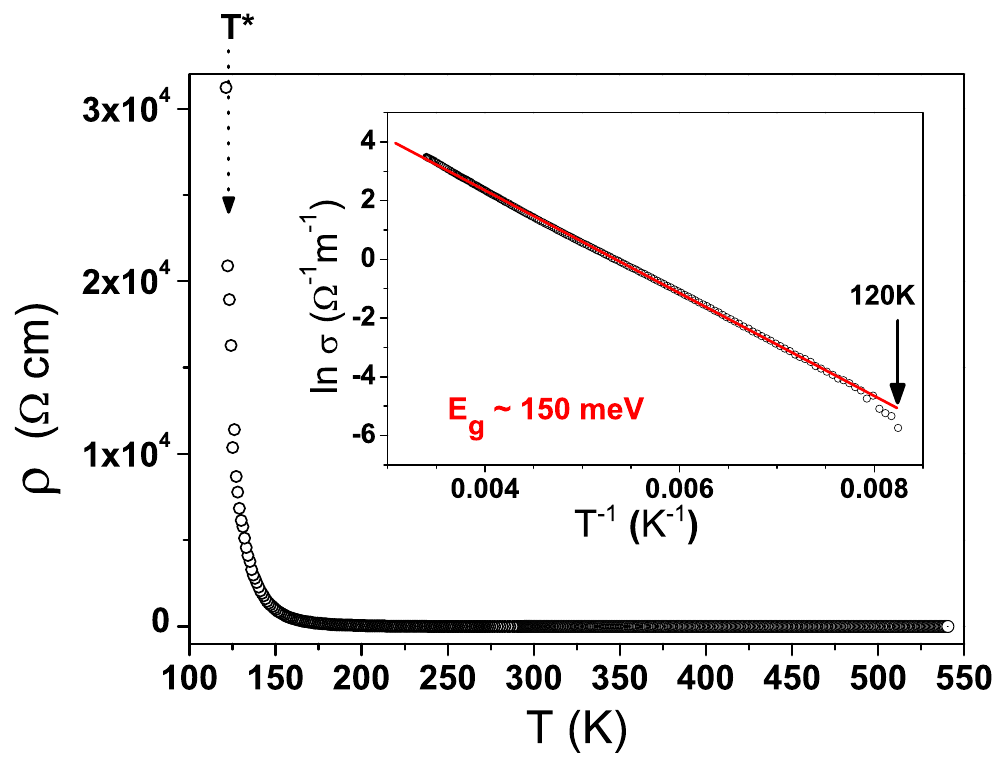}
\caption{Specific resistivity of In$_2$VO$_5$. The inset shows a
logarithmic plot of the conductivity versus 1/T below 300 K. (Colors
online)\label{spezrho}}
\end{figure}

The electrical resistivity $\rho$ of a pressed and sintered powder
pellet of In$_2$VO$_5$ in the temperature range of 100\,K to 540\,K
is shown in Fig.~\ref{spezrho}. In the high temperature regime
$\rho$ does not change significantly as indicated by $\rho$(290\,K)
= 3.30\,$\Omega$\,cm and $\rho$(500\,K) = 2.66\,$\Omega$\,cm whereas
below $\sim 150$\,K the resistivity increases dramatically. Except
at temperatures close to $T^*\sim 120$\,K the resistivity follows
the Arrhenius law which yields a linear dependence of the
logarithmic plot of the conductivity $\sigma = \rho^{-1}$ versus
$T^{-1}$ with the slope given by the activation energy $E_g\sim
150$\,meV (Fig.~\ref{spezrho}, inset). By approaching $T^*$ the
conductivity drops down indicating the transition from
semiconducting to insulating behavior. Notably, one can relate this
transition to the crossover in the $T$-dependence of the magnetic
susceptibility occurring around $T^*$ (Fig.~\ref{chi}). We recall
that at this temperature also the crystallographic $a$ and $b$ axes
begin to increase slightly (Fig.~\ref{lattice}), whereas the
anisotropic lattice contraction sets in below $T\sim 150$\,K
(Fig.~\ref{lattice}).

Standard optical characterization of In$_2$VO$_5$ reveals in the
infra-red region the expected vibration modes (phonons) below
800\,cm$^{-1}$\,$\approx$\,100\,meV (spectra not shown here). Above
1000\,cm$^{-1}$ transmittance is significantly reduced reaching a
constant value at $\sim 1100$\,cm$^{-1}$\,$\approx$\,135\,meV, which
corresponds approximately to the optical band gap and is in
agreement with the band gap $E_g$ obtained from the resistivity
measurements.

As suggested by the resistivity data insulating behavior is present below
$T^*\sim 120$\,K. Therefore, we studied the temperature dependence of
In$_2$VO$_5$ in the near infra-red - visible part of the electromagnetic
spectrum (4000 - 20000\,cm$^{-1}$). No changes in the absorbance were
detected and no indication of electronic $d-d$ transitions according to a
localized [VO$_6$]-chromophore were observed. This again corroborates the
picture of a semiconductor with a band gap (thermal or optical activation
energy) at lower energies than the studied spectral range, e.g. $<
0.5$\,eV.

\section{Magnetic resonance}

To elucidate the interplay between structure, conductivity and
magnetism in In$_2$VO$_5$ suggested by the experimental data in
Sections~\ref{crystal-structure}, \ref{magnetization} and
\ref{resistivity} we have studied the magnetic resonance of the
nuclear- and electronic spins at the vanadium site. NMR and ESR
probe on a local scale the electron-spin dynamics and spin
correlations.

\subsection{NMR measurements}

\begin{figure}
\includegraphics*[angle=0,width=\columnwidth]{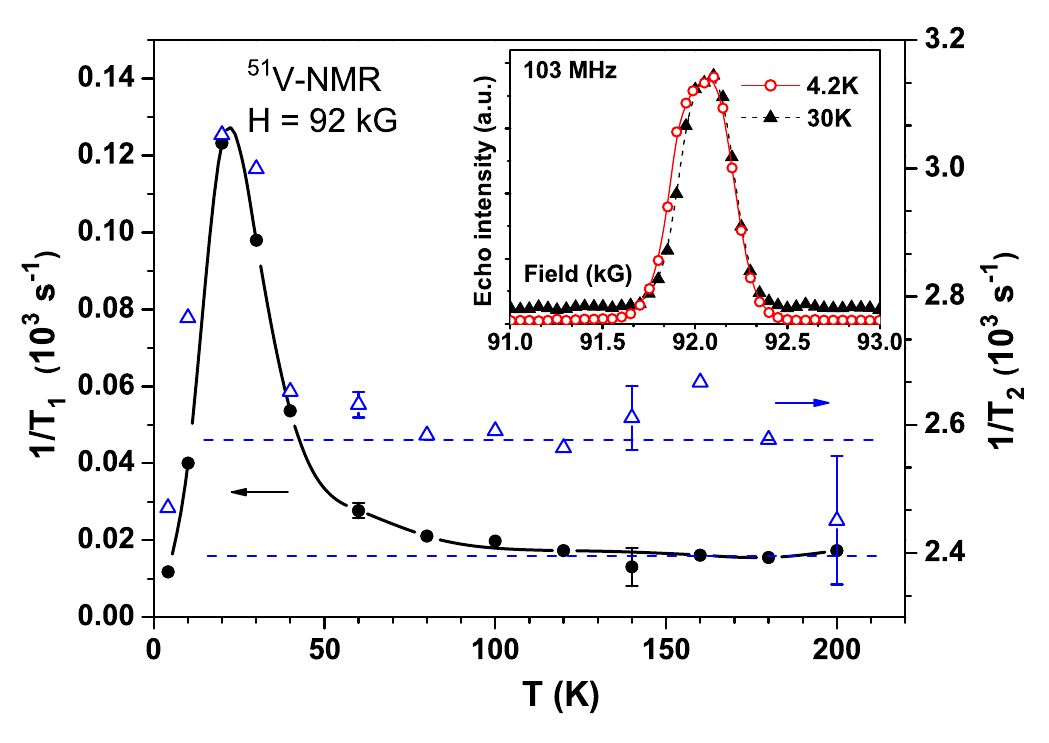}
\caption{$^{51}$V NMR signal at a frequency of 103 MHz at 4.2 and 30 K
(inset); Temperature dependence of 1/T$_1$ and 1/T$_2$ nuclear relaxation
rates measured at the maximum of the signal (main panel). (Colors
online)\label{nmr}}
\end{figure}

$^{51}$V NMR in \ivo\ has been studied at a frequency of 103\,MHz in
a temperature range 4.2 - 200\,K.  Typical spectra are shown in the
inset of Fig.~\ref{nmr}. The $^{51}$V nucleus has a spin $I=7/2$ and
possesses a nuclear quadrupole moment. Owing to the interaction of
the quadrupole moment with the gradient of the electrical crystal
field the $^{51}$V NMR spectrum may generally acquire a structure
consisting of the main central line and 6 satellites. However, in
the case of In$_2$VO$_5$ the satellites are unresolved in the powder
spectrum and its shape can be described by a single Gaussian line
profile (solid and dashed lines in the inset of Fig.~\ref{nmr}). The
NMR-lineshape, the resonance field and the linewidth are almost
unchanged in the studied temperature range. Only by approaching the
lowest temperature of 4.2 K can a small broadening of the signal and
a shift of its position to lower fields be observed. However, both
the nuclear transversal (spin-spin) and the longitudinal
(spin-lattice) relaxation rates, $1/T_2$ and $1/T_1$, show a
temperature dependence (Fig.~\ref{nmr}, main panel). There is a
sharp enhancement of both relaxation rates peaked at $\sim 20$ K
followed by their slowing down at higher temperatures. The 1/T$_2$
rate levels off at $\sim 70$\,K and exhibits a small hump at about
150\,K which only slightly exceeds the experimental error. $1/T_1$
continues to decay smoothly up to $\sim 100$\,K where it reaches a
constant value of 16\,s$^{-1}$.

Generally, the nuclear relaxation can be caused by electric
quadrupole and magnetic interactions. The strong enhancement of the
nuclear relaxation processes near 20\,K can be straightforwardly
related to the critical slowing down of magnetic fluctuations in the
electronic subsystem. The electron spins of vanadium, fluctuating
with some characteristic frequency $\omega _{sf}$, exert an
effective field at the vanadium nuclear spin owing to the on-site
hyperfine coupling. The transverse component of this field $h_{\rm
eff}$ contributes to the longitudinal nuclear spin relaxation as
$T_1^{-1}= \gamma_N^2h_{\rm eff}^2/[\omega _{sf}(1 +
\omega_{L}^2/\omega _{sf}^2)]$ \cite{Slichter}. Here $\gamma_N$ is
the nuclear gyromagnetic ratio and $\omega_L$ is the nuclear Larmor
precession frequency. The relaxation is maximum when the
fluctuations of the electron spins slow down to the NMR frequency,
$\omega_{\rm sf} = \omega_L$. The rate $1/T_1$ reduces in the case
of both fast- ($\omega_{\rm sf} > \omega_L$) and slow fluctuating
($\omega_{\rm sf} < \omega_L$) or static electron spins. The peak in
the $T$-dependence of $1/T_1$ thus clearly indicates a transition to
a magnetically ordered or very slowly fluctuating phase in
In$_2$VO$_5$ at $T_{\rm SRO}\sim 20$\,K. A similar spin fluctuating
relaxation mechanism may give a contribution $(1/T_2)_{\rm sf}$ to
the total transversal nuclear spin relaxation rate $1/T_2 =
1/T_2^\prime + (1/T_2)_{\rm sf}$ in addition to the contribution
$1/T_2^\prime$  arising from nuclear spin-spin couplings, e.g.
magnetic dipolar or indirect interactions \cite{Slichter}. The fact
that the peak in the dependence of $1/T_2$ versus $T$ occurs at the
same temperature $T_{\rm SRO}$ as for the $1/T_1$ rate corroborates
its assignment to the onset of a (quasi)-static magnetic order.

However, below $T_{\rm SRO}$ we do not observe a significant
broadening and/or splitting and shift of the NMR line indicative of
the development of the staggered magnetization in the long-range AFM
ordered state. The occurrence of the peak in the nuclear relaxation
rates is associated with only a small broadening and shift of the
NMR line at $T < T_{\rm SRO}$. It is therefore plausible that only a
short-range (glassy-like) magnetic order (SRO) is realized in
In$_2$VO$_5$ below $T_{\rm SRO}$ which can be detected by local
magnetic resonance techniques (see also Section \ref{esr} below) but
not by the bulk thermodynamic measurements discussed in Section
\ref{magnetization} and Ref.~\onlinecite{sh}.

The longitudinal relaxation rate 1/T$_1$ of the vanadium nuclei
arising due to the hyperfine coupling with the electron spins is
proportional to the imaginary part of the momentum $q$ and frequency
$\omega$ dependent electron dynamic susceptibility
$\chi^{\prime\prime} (q,\omega)$ at small energies as 1/T$_1 \propto
T\chi^{\prime\prime}(q,\omega)$ \cite{Moriya1962}. The independence
of 1/T$_1$ on temperature at $T > T^*$ implies then a Curie-like
dependence of $\chi^{\prime\prime}$ consistent with the behavior of
the susceptibility in the static limit (cf. Fig.~\ref{chi}).
Therefore, in spite of a significant reduction of the resistivity
and evidence for the delocalization of the $d$-electrons above $T^*$
from the ESR measurements (see Section~\ref{esr} below) In$_2$VO$_5$
has not yet reached the limit of a band metal for which a
$T$-independent susceptibility and a linear in temperature
Korringa-like relaxation 1/T$_1\sim T$  is expected
\cite{Korringa1950}.

Interestingly, the longitudinal relaxation rate 1/T$_1$ which probes
the dynamic spin susceptibility of the electronic system on the NMR
time scale and can also be sensitive to charge fluctuations via the
quadrupole relaxation channel, remains appreciably $T$-dependent
even far above $T_{\rm SRO}$. Thus, it is tempting to associate the
onset of this $T$-dependence with a crossover from band- to
localized AFM interacting $d$-states that is identified in the ESR
measurements (see Section \ref{esr} below) at the same
characteristic temperature $T^* \sim 120$\,K. Moreover, one can
speculate that a small hump in the $T$-dependence of the nuclear
spin-spin relaxation rate $1/T_2$ is related to the change of the
lattice contraction from the isotropic to anisotropic regime at
$\sim 150$\,K. Because these structural changes affect the
nearest-neighbor distances and bond angles between V sites, this
could alter the nuclear dipole-dipole interactions contributing to
$1/T_2^\prime$.

\subsection{ESR measurements}
\label{esr}

\begin{figure}
\includegraphics*[angle=0,width=\columnwidth]{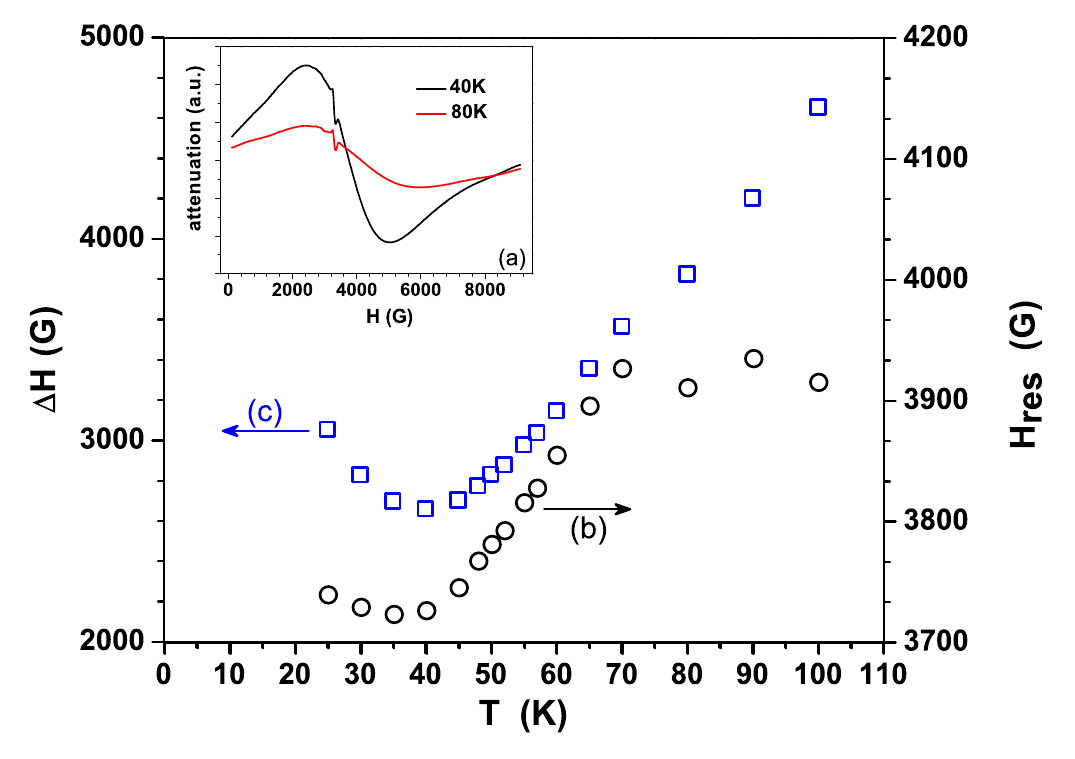}
\caption{(a) The field derivative of the ESR signal at X-band frequency of
9.5\,GHz is shown for two temperatures. The small sharp signal is due to
small amounts of ''isolated'' paramagnetic impurities. The temperature
dependence of the resonance field $H_{res}$ (b) and the line width $\Delta
H$ (c) of the main broad signal is given in the main panel. (Colors
online)\label{1ESR}}
\end{figure}

\begin{figure}
\includegraphics*[angle=0,width=\columnwidth]{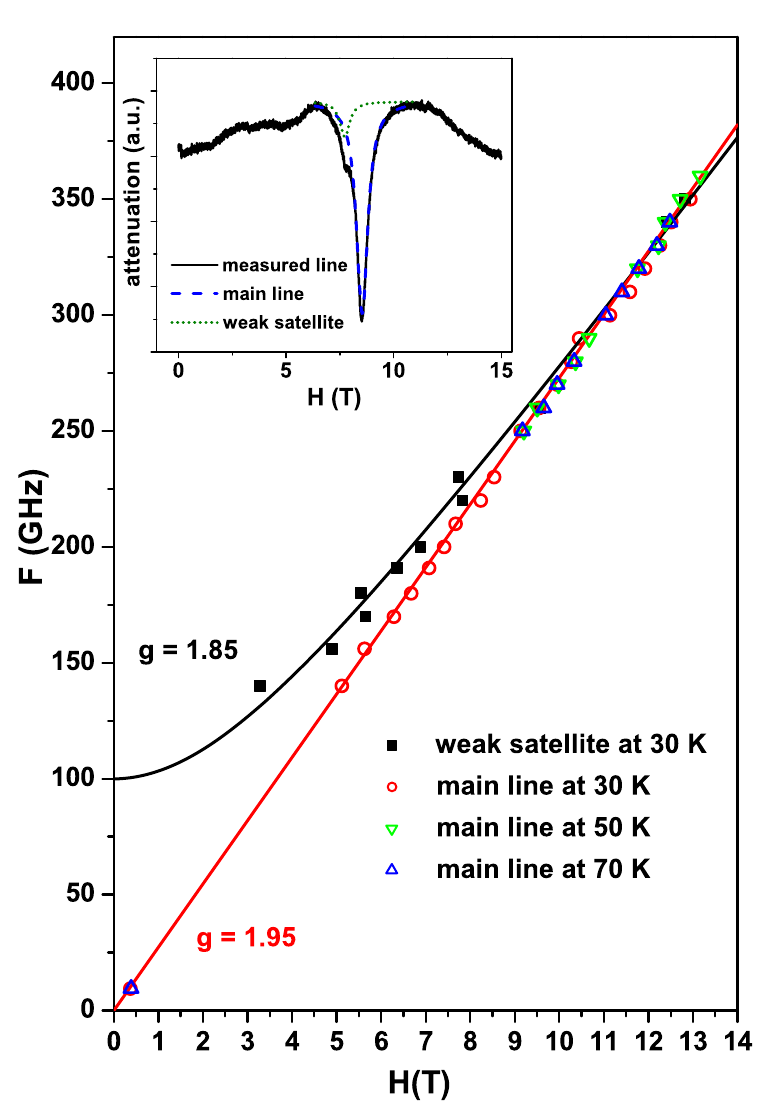}
\caption{A typical ESR spectrum at 230 GHz is shown in the inset. Two
resonance lines (denoted \emph{weak} and \emph{main}) are indicated within
the fitted signal. In the main panel the frequencies of these two
resonance lines are plotted versus the magnetic field for different
temperatures. (Colors online)\label{2ESR}}
\end{figure}

\begin{figure}
\includegraphics*[angle=0,width=\columnwidth]{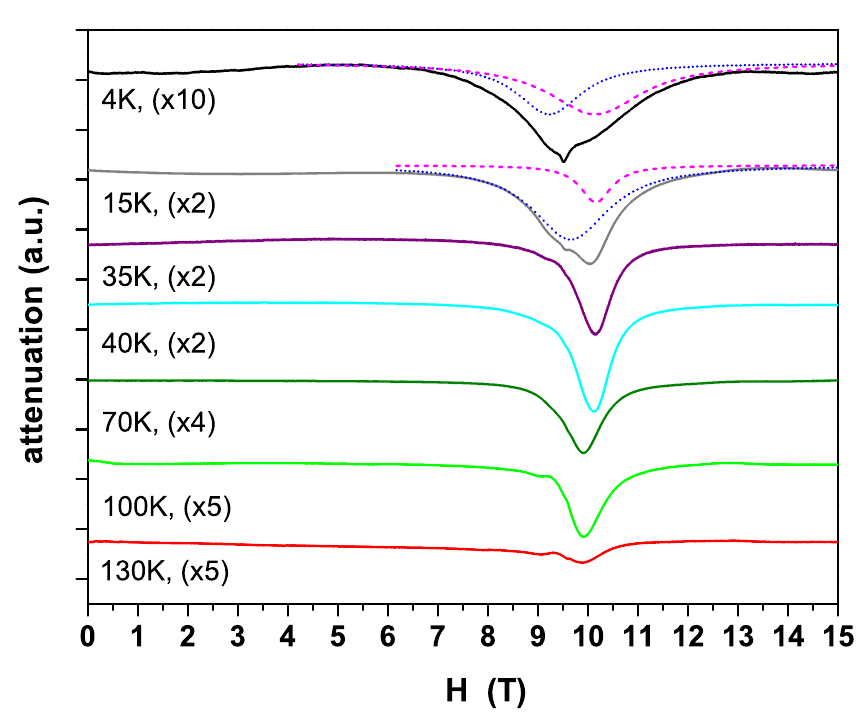}
\caption{Evolution of the ESR spectrum with temperature at $\nu =
270$\,GHz. (Colors online)\label{3ESR}}
\end{figure}

\begin{figure}
\includegraphics*[angle=0,width=\columnwidth]{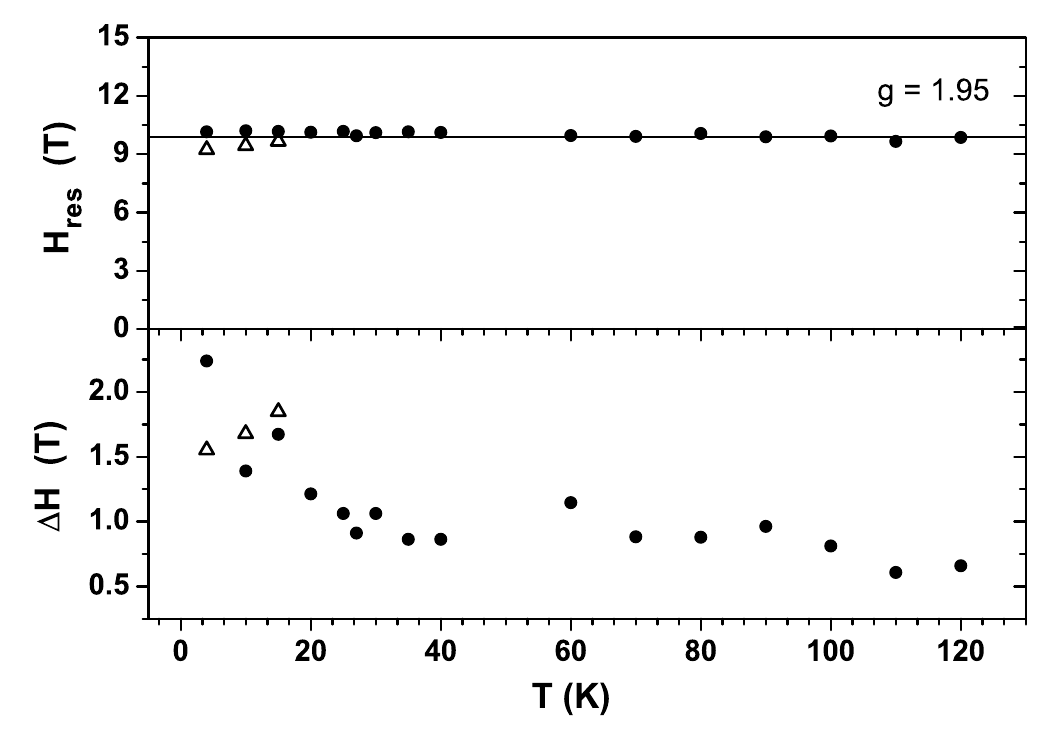}
\caption{Temperature dependence of $H_{res}$ and $\Delta H$ at $\nu =
270$\,GHz. Triangles correspond to the additional AFM mode developing at
$T\leq 15$\,K.\label{4ESR}}
\end{figure}

No ESR response from the powder sample of In$_2$VO$_5$ can be
detected at room temperature regardless the frequency $\nu$ of the
measurement. However, a well-defined ESR signal which can be
associated with the resonance of bulk V$^{4+}$ ($3d^1, S = 1/2$)
ions emerges below $\sim$ 130 K. At $\nu$ = 9.5 GHz (X-Band) ESR is
measured by a lock-in phase-sensitive detection with a sample placed
in the resonance cavity. With this technique the detected signal is
a field derivative of the microwave absorption $dP(H)/dH$. The shape
of the signal from In$_2$VO$_5$ can be described by a
single-derivative Lorentzian line profile  (Fig.~\ref{1ESR}a). The
line is very broad at high temperatures. However the width of the
signal $\Delta H$ rapidly decreases with decreasing the temperature
down to 40\,K and then starts to increase again (Fig.~\ref{1ESR}c).
The position of the resonance $H_{res}$ shifts to lower fields in
the same temperature range (Fig.~\ref{1ESR}b). Remarkably, the ESR
signal vanishes upon approaching a temperature of $\sim 20$\,K.

In the high frequency regime (140 - 360\,GHz) the transmission of the
microwave radiation through the sample has been measured as a function of
magnetic field without a resonance cavity. A typical spectrum at $\nu$ =
230 GHz is shown in the inset of Fig.~\ref{2ESR}. Unlike the X-Band
measurement, the fit of the absorption line requires two Lorentzians: A
strong main resonance line and a weak satellite on the left shoulder. The
frequency versus magnetic field relation of these two resonance modes is
shown in the main panel of Fig.~\ref{2ESR}. The $\nu (H)$ dependence of
the main signal is linear. Its extrapolation to low fields intersects the
origin and matches very well with the measurement at $\nu = 9.5$\,GHz. In
contrast to the main line, the $\nu (H)$-branch of the weak satellite has
an appreciable frequency offset $f_0 \sim 100$\,GHz corresponding to an
energy gap for this resonance excitation.

From the slope of the  main linear $\nu (H)$-branch the $g$-factor $g =
1.95$ can be calculated, in perfect agreement with the results of the
angular overlap model (AOM) calculations ($g_{\text{av}} = 1.94$ and the
principal $g$-values of 1.92, 1.94 and 1.96, see below). Note that at all
measured frequencies no deviation from the Lorentzian shape has been
observed, indicating that the anisotropy of the $g$-factor which should
yield the spread of the resonance fields in the polycrystalline samples is
smaller than the width of the resonance. With the linewidth of the order
of 1\,T  the estimated anisotropy of the $g$-factor is smaller than a few
percent which corroborates the AOM calculation.

The evolution of the ESR spectrum with temperature at $\nu =
270$\,GHz is shown in Fig.~\ref{3ESR}. Note that at this frequency
the weak satellite merges with the main line (Fig.~\ref{2ESR}).
Therefore, the signal can be described by a single Lorentzian
absorption profile in a wide temperature range. The $T$-dependence
of $H_{\rm res}$ and $\Delta H$ obtained from the Lorentzian fit is
shown in Fig.~\ref{4ESR}. $H_{res}$ is temperature independent
whereas $\Delta H$ continuously increases with lowering the
temperature. Below 20\,K the signal not only broadens but also
acquires an asymmetrical shape indicating the development of an
additional resonance mode. Fitting the line profile for $T\leq
15$\,K with two Lorentzians (Fig.~\ref{3ESR}) yields the
$T$-dependence of the linewidth and the resonance field for the
additional mode which is plotted in Fig.~\ref{4ESR}. One can observe
the shift and narrowing of this mode whereas the main absorption
line continues to broaden with $H_{\rm res} \approx$\,const.

The development of the additional mode corresponds nicely with the
occurrence of the SRO state in In$_2$VO$_5$ at $T_{\rm SRO}\sim
20$\,K suggested by the NMR data. It can be related to an AFM-like
collective magnetic excitation in the SRO regions which sharpens and
shifts away from the paramagnetic line. The coexistence of this mode
with the paramagnetic signal implies the presence of non-ordered
regions at $T<T_{SRO}$ which would be consistent with a continuous
spin-glass freezing scenario. Such an extremely broad ESR response
below $T_{SRO}$ can only be observed at high excitation frequencies
and not in a "low" frequency X-Band measurement.

Rapid disappearance of the ESR signal from the localized $d$-states
of V$^{4+}$ by heating the sample over a characteristic temperature
$T^* \sim 120$\,K is unusual. In particular, it coincides with a
significant increase of the conductivity of In$_2$VO$_5$ around this
temperature. Therefore, one can speculate that $T^*$ is a
characteristic temperature for delocalization of the $d$-electrons
which at $T > T^*$ begin to contribute to the conductivity. That
would naturally explain the vanishing of the ESR signal. Even if
delocalization occurs on a scale of some lattice constants it would
yield the momentum scattering of electrons which owing to the
significant spin-orbit coupling should result in extremely short
spin relaxation times.

The occurrence of the weak satellite mode with an excitation energy
gap $f_0 \sim 100$\,GHz is not expected if one considers the
resonance of an $S = 1/2$ system, like isolated V$^{4+}$ ions, or
ions coupled in a chain. Here one requires a correlated spin cluster
with a total spin $S_t\geq 1$. Feasible AFM correlations of V$^{4+}$
spins on a triangular pattern along the vanadium chain will be
discussed in Section~\ref{discussion} below. By considering, for
simplicity, an isolated AFM triangle of spins-1/2, one can
qualitatively explain the ESR-observations. Speculatively, the weak
satellite may correspond to a resonance transition between the
$|S_t=1/2; S^z=\pm1/2>$ ground state of the triangle and its high
energy $|S_t=3/2; S^z=\pm 1/2,\pm 3/2>$ state. In ESR the total spin
$S_t$ is a conserved quantity. Therefore, a transition between
different spin multiplets is forbidden. However, in the presence of
anisotropic magnetic interactions, such as e.g. the
Dzyaloshinskii-Moriya interaction, different spin states can be
mixed and the resonance between them may become visible\cite{sakai}.
In this scenario the excitation gap $f_0 \sim 100$\,GHz is a measure
of the energy separation between the $S_t=1/2$ and $S_t=3/2$ states
of the triangle.

\section{Discussion}
\label{discussion}

\begin{figure}
\begin{center}
\includegraphics*[angle=0, width=\columnwidth]{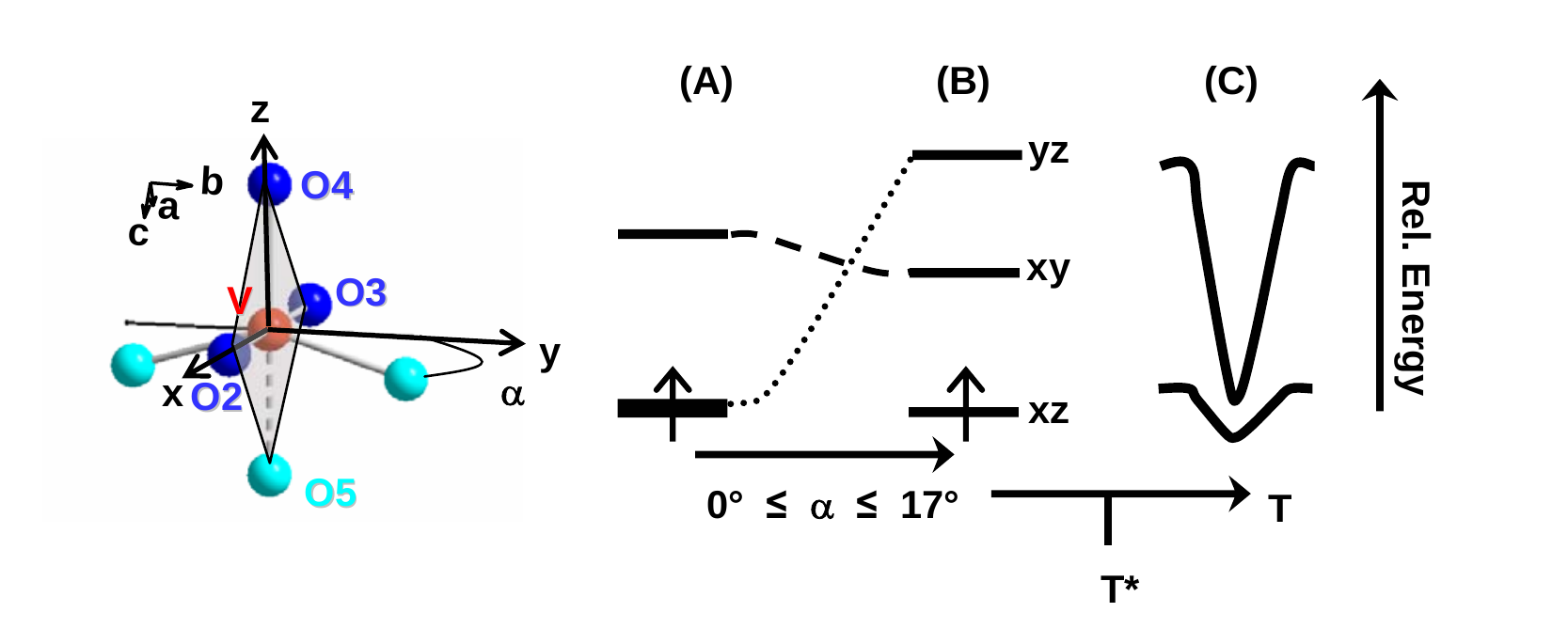}
\caption{The splitting of the t$_{2g}$-set for an "isolated"
[VO$_{5+1}$] ligand complex (3d$^1$ (S = 1/2) system) based on AOM
calculations is depicted within the crystal ($a, b, c$) and local
($x, y, z$) coordination frame. The grey shaded rectangle presents
the plane of the half-occupied d$_{xz}$ orbital. The relative energy
scales for the cases (A) and (B) are given as well as a sketch of
the proposed delocalized electronic structure (C). (Colors
online)\label{aom1}}
\end{center}
\end{figure}

In order to grasp the different aspects of the rich physics of
In$_2$VO$_5$ emerging from the structure, magnetization and
spectroscopy measurements it is useful to start with the properties
of the orbital states of the V$^{4+}$ ($3d^1$) ion. The degeneracy
of the ground state  $t_{2g}$-orbitals, where a single $d$-electron
resides will be lifted mainly by the $\pi$-antibonding interactions
with the O$^{2-}$ ligands for the [VO$_6$] metal-ligand complex. A
convenient method based on ligand field theory is an analysis within
the angular overlap model (AOM)\cite{AOM,jorgensen,CAMMAG}. The AOM
calculations include the interatomic distances and angles derived
from the single crystal structure determination, and therefore,
distortions from an octahedral ligand field can be accounted for.
Since the interaction between V$^{4+}$ and O$^{2-}$ depend on the
interatomic distances, it is quite reasonable to assume a first and
a second coordination sphere, [VO$_{5+1}$], comprising 5 nearest
neighbor oxygens O2, O3, O4 and 2$\times$O5 on the first one, and
the 6th neighbor O5 on the second sphere, respectively (see
Fig.~\ref{aom1} and Section~\ref{crystal-structure}). Then the
orbital splitting for all ligands within an orthogonal coordination
frame will correspond to sketch (A) in Fig.~\ref{aom1} with the
degenerate $d_{xz}$ and $d_{yz}$ orbitals being lowest in energy.
Inclusion of the observed deviation from linearity along [010] for
the two nearest O5 ligands parametrized by the angle $\alpha$
(Fig.~\ref{aom1}) will lift this degeneracy (Fig.~\ref{aom1}, sketch
(B)). The relative energy difference of the $d_{xy}$ to the lower
$d_{xz}$ orbital is then entirely dependent on the different amount
of the $\pi$-antibonding interaction for the first and second
coordination spheres. Since for the $d_{xz}$ orbital this
interaction, based upon distances, will be much smaller than the
former one, the ground state will be associated with the $d_{xz}$
orbital in any case for the local structure. To check these results
we have calculated the relative energy separation $\Delta E$ between
the $d_{xz}$ and $d_{xy}$ as a function of $\pi$-antibonding
interaction and the corresponding average molecular $g$-value
including spin-orbit-coupling. From these calculations one can
estimate that $\Delta E$ should be at least $\sim 0.25$\,eV to
account for $g\sim 1.95$ as observed in the ESR experiment. The
complete calculation including all distances and angle dependencies
of the ligands will shift the relative energy of the $d_{yz}$
orbital highest to $\sim 0.5$\,eV.

The ground state E(xz) configures the occupied $d$-orbital in a
$\delta$-stacking fashion along the $b$-axis which practically do
not overlap. On the other hand, the two excited states E(xy, yz) are
oriented in a $\pi$ arrangement in this direction, which might lead
to an effective overlap involving O5 orbitals whereby a significant
covalency is anticipated. In this scenario one would expect a
considerable degree of delocalization along the $b$ direction for
the electrons promoted into these overlapping molecular orbitals.
One may speculate about a narrow band formation that is depicted in
sketch (C), Fig.~\ref{aom1}, which would favor a ferromagnetic
polarization of the $d$-electrons. Strong indications for the
relevance of this scenario in the high temperature regime above
$T^*$ are given by significant conductivity, ferromagnetic sign of
the net exchange derived from the $\chi(T)$ measurements and the
absence of the ESR response which is suggestive of the itinerant
character of the $d$-states.

\begin{figure}
\includegraphics*[angle=0,width=\columnwidth]{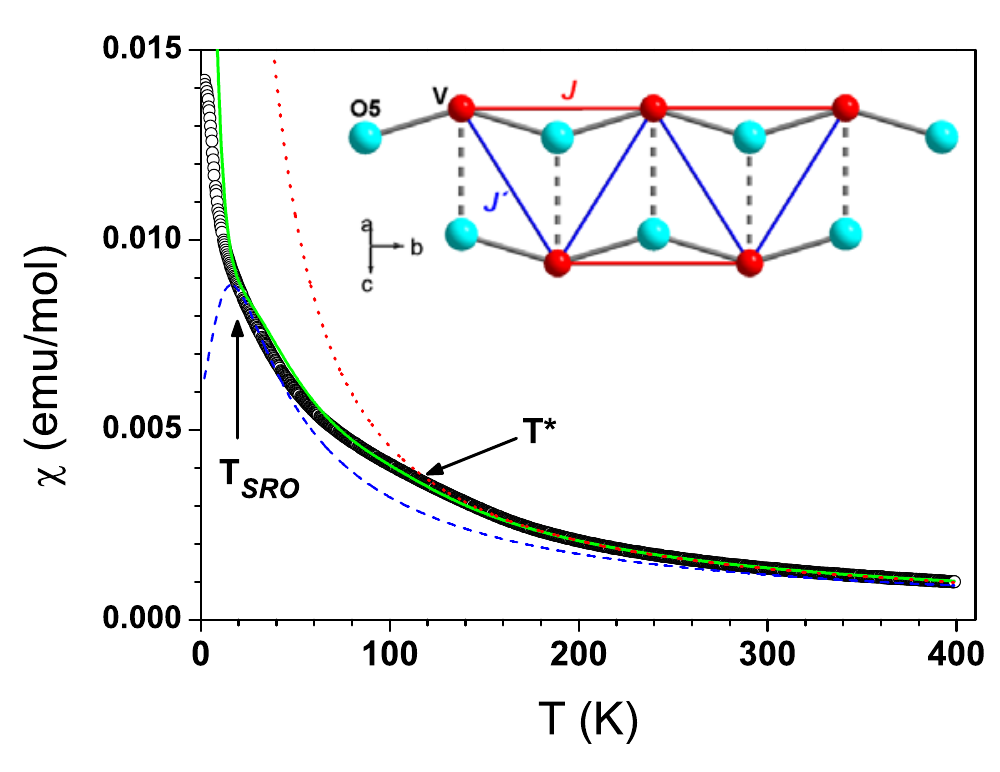}
\caption{Measured susceptibility of In$_2$VO$_5$ at 1\,T ($\circ$),
calculated susceptibilities for different models: HT series expansion
(red, dotted), Bonner-Fisher (blue, dashed) and a trimer (green), see
text. The inset gives a sketch of the zig-zag chain feature with the
interaction parameters $J$ and $J^{\prime}$. (Colors online)\label{mag}}
\end{figure}

Lowering the temperature yields a crossover to the anisotropic
thermal contraction which sets in at $\sim 150$\,K. Below $T^* \sim
120$\,K the $b$ lattice parameter even starts to increase slightly.
The shrinking of the $c$ lattice parameter implies a decrease of the
distance between the vanadium and the O5 ligand in the neighboring
chain (Fig.~\ref{mag}, inset) thus increasing the overlap of the
$xz$-orbitals with the $p$-orbitals of that ligand. Considering the
orientation of the $xz$-orbitals and the interchain bond angle
V\,-\,O5\,-\,V of 106$^\circ$ one would expect an increase of the
AFM superexchange between the vanadium chains. Concomitantly, owing
to the increasing $b$ lattice parameter, the width of the hybridized
band in the chain direction (Fig.~\ref{aom1}, sketch (C)) should
decrease and become thermally depopulated. One could speak therefore
of a temperature and structurally driven crossover from the
itinerant-like ferromagnetically (FM) correlated linear chain along
[010] with the effective exchange $J$ (FM) to the zig-zag chain of
localized spins with the AFM nearest neighbor interaction $J^\prime$
(AFM) (Fig.~\ref{mag}, inset).

Neglecting $J^\prime$, the best fit of the susceptibility data
$\chi(T)$ for $T > T^*$ according to a high temperature (HT) series
expansion\cite{baker} is obtained with $J \sim$ +22 K
(Fig.~\ref{mag}). A poor fit of the low temperature regime ($T <
T^*$) to a Bonner-Fisher model\cite{bonnerf} of an $S = 1/2$ chain
with $J^{\prime} \sim$ -25\,K is given as well in Fig. \ref{mag}.
Note that the corresponding result for a HT series
expansion\cite{baker} model is obtained for $J^{\prime} \sim
-12.5$\,K. Comparing these results gives an estimate of the ratio
$-J^{\prime}/J\sim 0.5$. Both limiting cases do not match the
experimental findings below $T^*$. The magnetic interactions seem to
change gradually in the regime between $T_{\rm SRO}\approx 20$\,K
and 120\,K.

To account for the situation occurring in the intermediate
temperature range we consider short-range spin correlations
extending over three neighbor vanadium sites in a triangle fashion
(Fig.~\ref{mag}, inset). For simplicity reasons we discuss an
isolated spin trimer as the smallest magnetically coupled entity.
The relative zero field energy of states of a trimer consisting of
spins $\textbf{S}_{A1}$ and $\textbf{S}_{A2}$ coupled together via
the exchange path $J$ and both of them coupled to the third spin
$\textbf{S}_{B}$ via $J^\prime$ can be deduced from the spin
Hamiltonian \cite{kahn}:

\begin{eqnarray}
\textbf{H} = -J^{\prime}(\textbf{S}_{A1}\cdot \textbf{S}_B +
\textbf{S}_{A2}\cdot
\textbf{S}_B)-J(\textbf{S}_{A1}\cdot \textbf{S}_{A2})\\
E(S,S^{\prime}) = -J^{\prime}[S(S+1)]/2 -
 (J-J^{\prime})[S^{\prime}(S^{\prime}+1)]/2
\end{eqnarray}

with \textbf{S}$^{\prime}$ = \textbf{S}$_{A1}$+\textbf{S}$_{A2}$ and \textbf{S} = \textbf{S}$^{\prime}$ + \textbf{S}$_B$.\\

\begin{figure}
\includegraphics*[angle=0,width=\columnwidth]{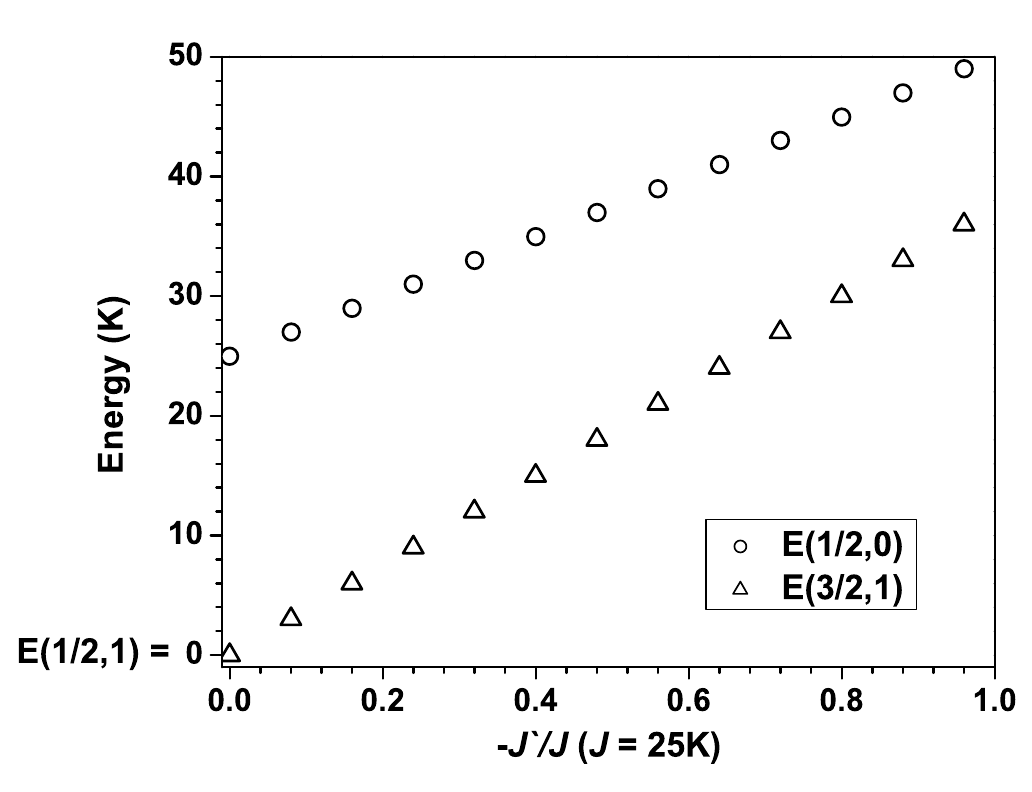}
\caption{Dependence of the relative energy of states on $J^{\prime}$ for a
distorted triangle with the E(1/2,1) set as the origin.\label{energyl}}
\end{figure}

The relative energy of states $E(S,S^{\prime})$ in zero field is
depicted in Fig.~\ref{energyl}. If we use the observed isotropic
local $g$-tensor, the susceptibility can be derived by applying the
Van-Vleck formula after adding the Zeeman perturbation. Since the
interaction parameter $|J^{\prime}|$ is expected to increase as the
temperature decreases, we let, for the interpretation within this
model, $J^{\prime}$ be a function of temperature: $J^{\prime}(T) =
-14/[1+\text{exp}((T-90)/14)]$ and $J = 25$\,K. The result for a
trimer is given in Fig.~\ref{mag} and qualitatively matches the
experimental data above $T_{SRO} \approx 20$\,K, where NMR and ESR
indicate a drastic slowing down of spin fluctuations.

Despite obvious oversimplifications the principal features of the
susceptibility data obtained for In$_2$VO$_5$ and the observation of
the additional gapped mode in the ESR experiment are reasonably
explained by assuming the model of a magnetic triangle above 20\,K.
The energy gap between the E(1/2,1) ground and the first excited
state E(3/2,1) corresponds to $\sim$ 15 K ($-J^{\prime}/J = 14/25 =
0.56$) for an isolated triangle, see also Fig.~\ref{energyl}. From
ESR measurements this gap $f_0 \sim 5$\,K is somewhat smaller,
possibly because of the obvious correlations between the magnetic
units (triangles), which are arranged by inversion symmetry in an
anti-parallel alignment along the $b$-axis and thereby form the
zig-zag chain. The large Curie constant of 0.615 between $\sim
120$\,K and 20\,K might indicate substantial
contributions/admixtures of the almost degenerate states E(1/2,1)
and E(3/2,1) for a triangle that are even more enhanced for the
zig-zag chain. This might be one reason that the nominally
"forbidden" resonance transition between these two states is
observed in the ESR experiment.

\section{Summary}

In summary, from measurements of structural parameters,
magnetization, electrical resistivity and magnetic resonance of
In$_2$VO$_5$ we have presented extensive experimental evidence for
the interplay of crystal structural and electronic structure changes
that induce a transition from a semiconducting behavior with
ferromagnetic correlations at high temperatures to the insulating
regime with predominantly antiferromagnetic interactions between
vanadium localized moments at low temperatures. A simple model of
correlated spins coupled along the $b$ crystallographic axis on a
triangle pattern qualitatively explains the main observations. Our
results bring new insights into the rich physics of complex vanadium
oxides and call for theoretical modeling of the band structure and
magnetic interactions in In$_2$VO$_5$. First results recently
published by Schwingenschl\"ogl \cite{Schwing} are in line with our
experimental findings above $T^*$.

\begin{acknowledgments}
This work was supported by the DFG thorough SFB 608. The authors would
like to thank D. Senff and M. Braden for using the low temperature powder
x-ray diffraction equipment and N. Tristan for technical support. Valuable
discussions with D. Khomskii are gratefully acknowledged. The work of EV
was supported by the DFG grant BU 887/5-1 and by the Russian Foundation
for Basic Research through grant No. 07-02-01184-a. MY acknowledges the
Fellowship from the DAAD.
\end{acknowledgments}






\bibliography{InVO_v13}

\end{document}